\documentclass[]{Liebert_Author}
\usepackage{multicol}
\usepackage{setspace}
\usepackage{titlesec}
\usepackage{orcidlink} 
\usepackage{float}
\usepackage{lettrine}

\titleformat{\section}
  {\sffamily\normalsize\bfseries} 
  {\thesection.}                 
  {0.5em}                        
  {}

\titleformat{\subsection}
  {\sffamily\normalsize\itshape} 
  {\thesubsection.}              
  {0.5em}
  {}

\titleformat{\subsubsection}
  {\sffamily\normalsize}         
  {\thesubsubsection.}           
  {0.5em}
  {}

\date{}
\title{Fluorescent Biomolecules Detectable in Near-Surface Ice on Europa} 

\author{
    Gideon Yoffe\orcidlink{0000-0002-1451-6492}\thanks{Department of Earth and Planetary Sciences, Weizmann Institute of Science, Rehovot, 76100, Israel.}, 
    Keren Duer-Milner\orcidlink{0000-0003-4428-2700}\thanks{Leiden Observatory, Sterrenwachtlaan 11, 2311 GW Leiden, Netherlands.}, 
    Tom Andre Nordheim\orcidlink{0000-0001-5888-4636}\thanks{Johns Hopkins University Applied Physics Laboratory, 11100 Johns Hopkins Rd, Laurel, MD 20723, United States.}, 
    Itay Halevy\orcidlink{0000-0002-7325-8139}\footnotemark[1], 
    Yohai Kaspi\orcidlink{000-0003-4089-0020}\footnotemark[1]
}

\begin{document} 

\maketitle 


\begin{abstract}
Europa, Jupiter's second Galilean moon, is believed to host a subsurface ocean in contact with a rocky mantle, where hydrothermal activity may drive the synthesis of organic molecules. Of these molecules, abiotic synthesis of aromatic amino acids is unlikely, and their detection on Europa could be considered a biosignature. Fluorescence from aromatic amino acids, with characteristic emissions in the  200-400 nanometer wavelength range, can be induced by a laser and may be detectable where ocean material has been relatively recently emplaced on Europa's surface, as indicated by geologically young terrain and surface features. However, surface bombardment by charged particles from the Jovian magnetosphere and solar ultraviolet (UV) radiation degrades organic molecules, limiting their longevity. We model radiolysis and photolysis of aromatic amino acids embedded in ice, showing dependencies on hemispheric and latitudinal patterns of charged particle bombardment and ice phase. We demonstrate that biosignatures contained within freshly deposited ice in high-latitude regions on the surface of Europa are detectable using laser-induced UV fluorescence, even from an orbiting spacecraft. Key Words: Europa---(moon)---Amino acids---Spectroscopy.

\end{abstract}

\begin{multicols}{2}

\section{Introduction} \label{introduction}

\lettrine[lines=2]{E}{uropa} is a prime candidate in the search for extraterrestrial life due to its subsurface ocean, believed to be in contact with a rocky mantle \citep{carr1998evidence}. 
Hydrothermal activity on the ocean floor, driven by tidal heating, could support serpentinization, generating hydrogen and synthesizing organic molecules from inorganic precursors \citep{zolotov2001composition}. 
The generation of hydrogen through serpentinization is particularly significant because it provides a crucial ingredient for the synthesis of organic molecules \citep{wang2014serpentinization}, potentially fueling a subsurface biosphere. These processes mirror early Earth conditions where hydrothermal systems may have been a potential setting for the origin of life by providing energy and essential building blocks for complex organic chemistry \citep{martin2008hydrothermal}.

Aromatic amino acids, namely phenylalanine, tyrosine, and tryptophan, are particularly intriguing in the search for extraterrestrial life. These molecules are essential for life as they play critical roles in cellular processes, including protein synthesis and enzyme function \citep{pittard2008biosynthesis}. Their unique structures, featuring benzene or indole rings, enable them to absorb UV light and fluoresce distinctly in the 200-400 nm range \citep{beaven1952ultraviolet}. 
This fluorescence makes them valuable as unique tracers in fields like medicine and biochemistry \citep{yamashita2003chemical}, as well as within the context of the origin of- and the search for- life \citep{ehrenfreund2006experimentally}.
The fluorescence of aromatic amino acids is utilized for their detection in aquatic and icy settings on Earth \citep{eshelman2019watson}, as well in planetary contexts, where, for example, the SHERLOC instrument \citep{beegle2015sherloc} aboard the Perseverance rover on Mars utilizes deep ultraviolet (DUV) laser-induced fluorescence to detect and analyze organic compounds and minerals on the Martian surface \citep{bhartia2021perseverance}. Similar conceptual frameworks have been proposed for more extreme settings of solar system exploration, such as in-situ or remote laser-induced spectroscopic analysis of meteorites \citep{lymer2021uv}.


Simple amino acids such as glycine and alanine have been detected in various abiotic settings, including carbonaceous chondritic meteorites \citep{pizzarello1991isotopic} and hydrothermal environments \citep{zhang2017prebiotic}, where the Strecker synthesis is a well-established generation pathway \citep{koga2022synthesis}. In contrast, the synthesis of aromatic amino acids, with their more complex molecular structures, typically relies on multi-enzymatic metabolic pathways, such as the Shikimate pathway, which converts simple carbohydrates into aromatic compounds \citep{maeda2012shikimate}. The rarity of non-biological pathways for producing aromatic amino acids makes them compelling biosignatures \citep{georgiou2018functional}.

The recent detection of 14 out of the 20 proteinogenic amino acids, including tentative evidence for aromatic amino acids, in samples returned from asteroid Bennu expands the possibilities of prebiotic chemistry that occurred in the protosolar disk \citep{Glavin2025}. The presence of these compounds, formed through a combination of photochemical, radiolytic, and aqueous processes in ammonia-rich conditions, suggests that complex organic synthesis can take place in the early Solar System, broadening our understanding of the chemical precursors available for the emergence of life. Notably, polycyclic aromatic hydrocarbons (PAHs), which are structurally complex organic molecules composed of multiple aromatic rings, can also form abiotically through photochemical processing on surfaces such as water ice or silicate grains in stellar systems and protoplanetary disks \citep{kahan2007photolysis, henning2013chemistry, noble2020influence}. PAHs exhibit fluorescence similar to aromatic amino acids and produce distinct near- and mid-infrared (NIR and MIR) emission features \citep{aihara1992aromatic, peeters2021spectroscopic}, making them valuable tracers of organic chemistry in extraterrestrial environments \citep[e.g.,][]{giese2022experimental, xu2024organic}. Their presence, alongside amino acids in Bennu’s pristine material, underscores the diversity of prebiotic organic molecules that could have been available during planetary formation and the early evolution of life.  


On Earth, a marginal detection of tryptophan in the Lost City hydrothermal field represents the only known case where the synthesis of aromatic amino acids has been suggested to occur through an abiotic process \citep{menez2018abiotic}. Barring contamination from biological sources, this discovery is significant, as similar hydrothermal vent systems may exist on Europa. These environments could offer conditions conducive to the synthesis of complex organic molecules via both biotic and previously unexplored abiotic mechanisms in the extant Solar System \citep{lang2020habitability}.


Recently, a novel probabilistic approach introduced the molecular assembly index (MA), a measure that quantifies the assembly complexity of molecules \citep{marshall2021identifying}. According to this measure, molecules with an MA score greater than 15 are extremely unlikely to form abiotically. Tyrosine and tryptophan have MA scores of 10 and 12, respectively, as opposed to glycine, for example, which has an MA score of 4. Thus, while the abiotic synthesis of the aromatic species is possible -- it is not likely, which further elucidates the dearth of discovery of these species in abiotic settings.
On Earth, the concentration of tryptophan has been measured to be about 0.1 parts-per-billion (ppb) in the barren regions of the well-mixed ocean, where its synthesis is least likely, \citep{yamashita2004chemical}. The average abundances of tyrosine and phenylalanine are about 36 and 66 times higher, respectively \citep{moura2013relative}. While the observed concentrations cannot be directly tied to specific synthesis mechanisms, we consider them a conservative estimate for the initial concentration of aromatic amino acids in a well-mixed Europan ocean that contains putative life.

The detection of salts and ocean-derived compounds on Europa's surface suggests that subsurface material may be reaching the surface \citep{zolotov2001composition}. Recent JWST observations of carbon dioxide, its spatial distribution pattern, and isotope ratios also indicate an internal source of carbon \citep{trumbo2023distribution, villanueva2023endogenous}, further supporting this hypothesis. 

Several mechanisms could transport material from Europa's subsurface ocean to its surface. Cryovolcanism can deposit ocean material on the surface \citep{sparks2017active}, while tectonic activity can create pathways for subsurface material to ascend \citep{cashion2024europa}. Diapirism may also transport ocean material upward with buoyant ice \citep{pappalardo2004origin}.
Of these mechanisms, plumes are especially intriguing. Observations from the Hubble Space Telescope and Galileo spacecraft suggest water vapor plumes erupting from Europa’s surface, potentially depositing oceanic materials onto the ice \citep{roth2014transient}-- a process that is more frequent and better understood on Enceladus \citep{postberg2011salt}. This mechanism of emplacing ocean material onto the surface bypasses many complexities associated with other mechanisms, whose efficiency and timescales are not well constrained, and may involve significant alteration or mixing of the transported material during its ascent \citep{travis2012whole}.

Lastly, the detectability of these molecules is influenced not only by the rates of their delivery to the surface but, importantly, by their longevity in the harsh surface environment. 
Intense radiation due to electrons and ions from Jupiter’s magnetosphere \citep{nordheim2018preservation, nordheim2022magnetospheric}, whose penetration depth depends primarily on the particles' kinetic energy and can range between nanometers up to nearly a meter deep, and solar UV radiation, whose penetration depth varies similarly by many orders of magnitude -- from millimeters to meters -- depending on the optical properties of the ice \citep{orzechowska2007ultraviolet, johnson2012ultraviolet}. These, in turn, depend on the ice's thermal state and history \citep{he2022refractive}, irradiation intensity \citep{strazzulla1992ion}, and additional processes that are stochastic on short time-scales ($<$10$^3$ years), such as impact gardening \citep{costello2021impact}.
We modeled the radiolytic and photolytic degradation mechanisms independently and subsequently combined them to estimate the integrated longevity of aromatic amino acids embedded in Europan near-surface ice.

\section{Modeling Degradation Mechanisms} \label{degradation_mechanisms}

\subsection{Radiolysis} \label{radiolysis}

Studies of the Jovian magnetosphere, specifically at Europa's orbital distance \citep{mauk2004energetic, nordheim2022magnetospheric}, highlighted key aspects of the charged particles impacting Europa, specifically that: (1) Charged particles span kinetic energies from approximately 10 KeV to 100 MeV \citep{nordheim2018preservation}, where high kinetic-energy particles are much sparser than low kinetic-energy particles. (2) Major ions bombard Europa’s surface relatively uniformly \citep{nordheim2022magnetospheric}. (3) Energetic electrons exhibit a strong latitude- and hemisphere-dependent pattern. Electrons with kinetic energies below $\sim$20 MeV drift faster than Europa’s orbital velocity, impacting the trailing hemisphere. More energetic electrons, which are scarcer, impact the leading hemisphere because they have net retrograde drifts relative to Europa's orbital motion \citep{nordheim2018preservation}.
As a result, electron bombardment patterns form radiation ``lenses'' focused on the equatorial regions, extending to the mid-latitudes \citep{nordheim2018preservation}.

We implemented a simplified approach to simulate the interaction of charged particles with near-surface ice, utilizing methods similar to those used in previous studies \citep{nordheim2018preservation} (see Appendix \ref{methods_radiolysis}). Our model accounts for electrons and the three most abundant magnetospheric ions near Europa’s orbit: H$^{+}$, O$^{2+}$, and S$^{3+}$ \citep{nordheim2022magnetospheric}. Although there is some uncertainty regarding the charge states and their distribution among magnetospheric ions, our selection of these ions is primarily based on observational data \citep{mauk2004energetic}.
To simplify the simulation process, we assumed a uniform bombardment pattern for the ions. While additional factors, such as an ocean-induced magnetic field \citep{kivelson2004magnetospheric}, could be included, it has been shown that their impact on the bombardment pattern is relatively minor \citep{nordheim2022magnetospheric}, while adding significant complexity to the simulation. Moreover, deviations from uniform ion bombardment are most pronounced in low-latitude regions where electron energy deposition peaks, and thus the effect of this non-uniformity is small when considering the total surface dose from all charged particles \citep{nordheim2022magnetospheric}.
The surface was modeled as pure water ice with a density of 0.5 g cm$^{-3}$, incorporating the effects of porosity. This density is consistent with an expected porosity of approximately 0.3-0.5 for vapor-deposited water ice under Europan conditions \citep{mitchell2017porosity}.

To compute the longevity of amino acids under charged particle bombardment on Europa, we use a radiolytic constant of 0.034 MGy$^{-1}$
for a mixture of tryptophan, phenylalanine, and tyrosine, based on experimental values and scaled radiolytic reaction rates \citep{cataldo2011solid, gerakines2012situ} (see Appendix \ref{methods_radiolysis}).
The radiolytic constant can exhibit considerable variability that depends both on surface conditions and the distribution of the organic material \citep{gerakines2012situ, pavlov2024radiolytic}.
For example, hydrated salts on Europa's surface, such as magnesium sulfate (MgSO$_4$$\times$$n$H$_2$O), can absorb radiation, reducing the formation of radiolytic products and stabilizing amino acids against degradation over geological timescales \citep{carlson1999sulfuric, brown2013salts}. 
Additionally, recent experiments demonstrated that amino acids within organic matter degrade much slower than free amino acids in ice, suggesting their longevity may be significantly extended under realistic Europan conditions \citep{pavlov2024radiolytic}. 

Electrons bombarding the trailing hemisphere are considerably more numerous but less energetic ($\gtrsim20$ MeV), resulting in the deposition rate of more than an order of magnitude more energy into the ice. Furthermore, these less energetic electrons exhibit smaller penetration depths, leading to increased radiolytic degradation of amino acids within the upper layers.
Although energetic ions are sparser, they are also effective at degrading amino acids in the near-surface ice, because the two heavier ions deposit most of their energy within the top layer of the near-surface ice, in contrast to electrons, which tend to penetrate deeper (a detailed discussion can be found in Appendix \ref{methods_radiolysis}).

\subsection{Photolysis\label{photolysis}}

The photolytic degradation of amino acids can be approximated as a first-order exponential decay process influenced by wavelength-dependent photolytic reaction rates \citep{johnson2012ultraviolet} (see Appendix \ref{methods_photolysis}).
We considered the spectral range for photolysis to be between 147 and 342 nm, where the lower limit of 147 nm was chosen based on experimental evidence indicating significant attenuation of photons by water ice below this wavelength \citep{warren2019optical, he2022refractive}, while the upper limit corresponds to the dissociation energy of the C$_\alpha$-C$_\beta$ bond in glycine (about 3.62 eV \citep{luo2002handbook}) (see Appendix \ref{methods_photolysis}). 
Reaction rates for phenylalanine were extrapolated from measured values under equatorial surface conditions on Europa \citep{johnson2012ultraviolet}, and the rates for tyrosine and tryptophan were adjusted to reflect differences caused by molecular structure, as observed in $\gamma$-ray radiolysis experiments \citep{cataldo2011solid} (a detailed discussion about the photolytic spectral range and reaction rates can be found in Appendix \ref{methods_photolysis}). 
The annual mean solar flux incident on Europa’s surface was calculated by accounting for the latitudinal dependence, Europa's time in Jupiter’s shadow and its disk-averaged albedo, and the attenuation of this radiation with depth in the ice was modeled using a Beer-Lambert decay law, governed by ice extinction (see Appendix \ref{methods_photolysis}).

Amorphous solid water (ASW) is considered a characteristic phase of near-surface ice on Europa, especially in high-latitude regions \citep{hansen2004amorphous}, maintained by mechanisms such as impact gardening \citep{mastrapa2013amorphous} and slow crystallization time scales \citep{ mitchell2017porosity}.
Recently, wavelength-dependent extinction coefficient profiles of vapor-deposited ice, in both amorphous and crystalline forms, were measured under different deposition temperatures and shown to exhibit significant backscattering in the UV-visible range \citep{he2022refractive}. 
Additionally, it was shown that post-deposition heating changes the optical properties of amorphous vapor-deposited ice, often increasing the extinction coefficient by more than an order of magnitude \citep{he2022refractive}.
This is significant because photolytic degradation of amino acids by solar UV photons is potent when they are embedded in polycrystalline ice, which is nearly transparent at UV-visible wavelengths \citep{warren2019optical}, and results in very short half-lives of amino acids under surface conditions found on Europa \citep{johnson2012ultraviolet}. 
The phase of the ice on Europa is predominantly determined by crystallization and amorphization effects, which depend in turn on temperature and the deposited dose by charged particles \citep{strazzulla1992ion, mitchell2017porosity}. At the higher latitudes ($>$ 60$^\circ$), where surface temperatures are $<$ 110 K \citep{ashkenazy2019surface}, the crystallization rate is slow ($>$ 10$^5$ years) \citep{mitchell2017porosity}, causing ASW to remain amorphous over longer time scales \citep{berdis2020europa}. 
In contrast, at mid-latitudes, where surface temperatures are higher, free radicals that form by irradiation are more mobile and likely to recombine with the crystalline structure, thus reducing the effectivity of amorphization through irradiation \citep{strazzulla1992ion}.

We leveraged these measured wavelength-dependent extinction coefficients of vapor-deposited ice \citep{he2022refractive} to compute the attenuation of solar UV radiation that penetrates near-surface ice of similar characteristics.
For each location on Europa's surface, the initial extinction coefficient profile for the near-surface ice column was determined by the mean surface temperature at that location \citep{ashkenazy2019surface}.
Then, we accounted for the change in the extinction coefficient profile caused by diurnal-scale post-depositional heating by interpolating the measured change between the extinction coefficient profile at the deposition temperature (i.e., average diurnal temperature) and the maximal diurnal temperature \citep{he2022refractive} (see Appendix \ref{methods_ice}).
Finally, we evolved the extinction coefficient profiles through time, as a function of depth, by considering the temperature- and deposited dose-driven crystallization and amorphization processes \citep{strazzulla1992ion, mitchell2017porosity} (see Appendix \ref{app_amorphization_crystallization}), and derived photolytic half-lives of aromatic amino acids embedded therein.

\begin{figure*}[t!]
\centering
\rotatebox[origin=c]{0}{\includegraphics[scale = 0.58]{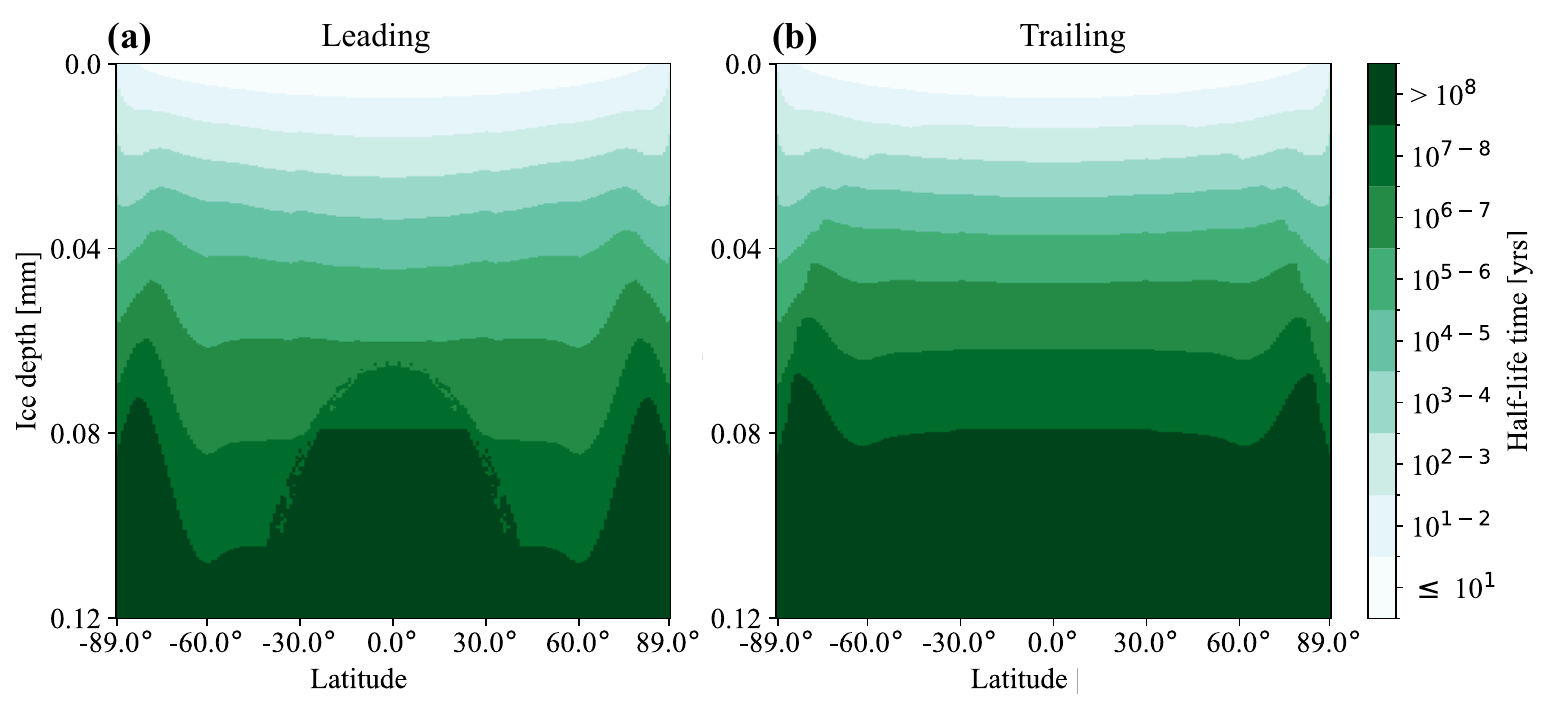}}
  \caption{Photolytic half-life times of aromatic amino acids embedded in thermally-evolved vapor-deposited ice that was subject to amorphization and crystallization processes as a function of latitude, depth, and time. \textbf{(a)} Leading hemisphere. \textbf{(b)} Trailing hemisphere.}
     \label{Fig_photolysis}
\end{figure*}

Fig. \ref{Fig_photolysis} shows the derived photolytic half-lives of aromatic amino acids as a function of latitude and depth for the trailing and leading hemispheres. The longest considered half-life time indicates the slowest estimate for the mean turnover rate of Europa's surface ($\sim$10$^8$ years -- \cite{doggett2009geologic}).
The apparent difference between the two hemispheres, particularly around latitude 50$^\circ$, can be explained by a considerably faster amorphization of ice in those regions at the trailing hemisphere, a process that increases ice opacity (see Appendix \ref{app_amorphization_crystallization}).
The difference in amorphization rates is driven, in turn, by lens-like bombardment patterns of energetic electrons that differ between the hemispheres, resulting in higher dose deposition in the trailing hemisphere, particularly in the mid-latitudes \citep{nordheim2018preservation} (see Appendix \ref{app_amorphization_crystallization}).

\begin{figure*}[t!]
\centering
\rotatebox[origin=c]{0}{\includegraphics[scale = 0.67]{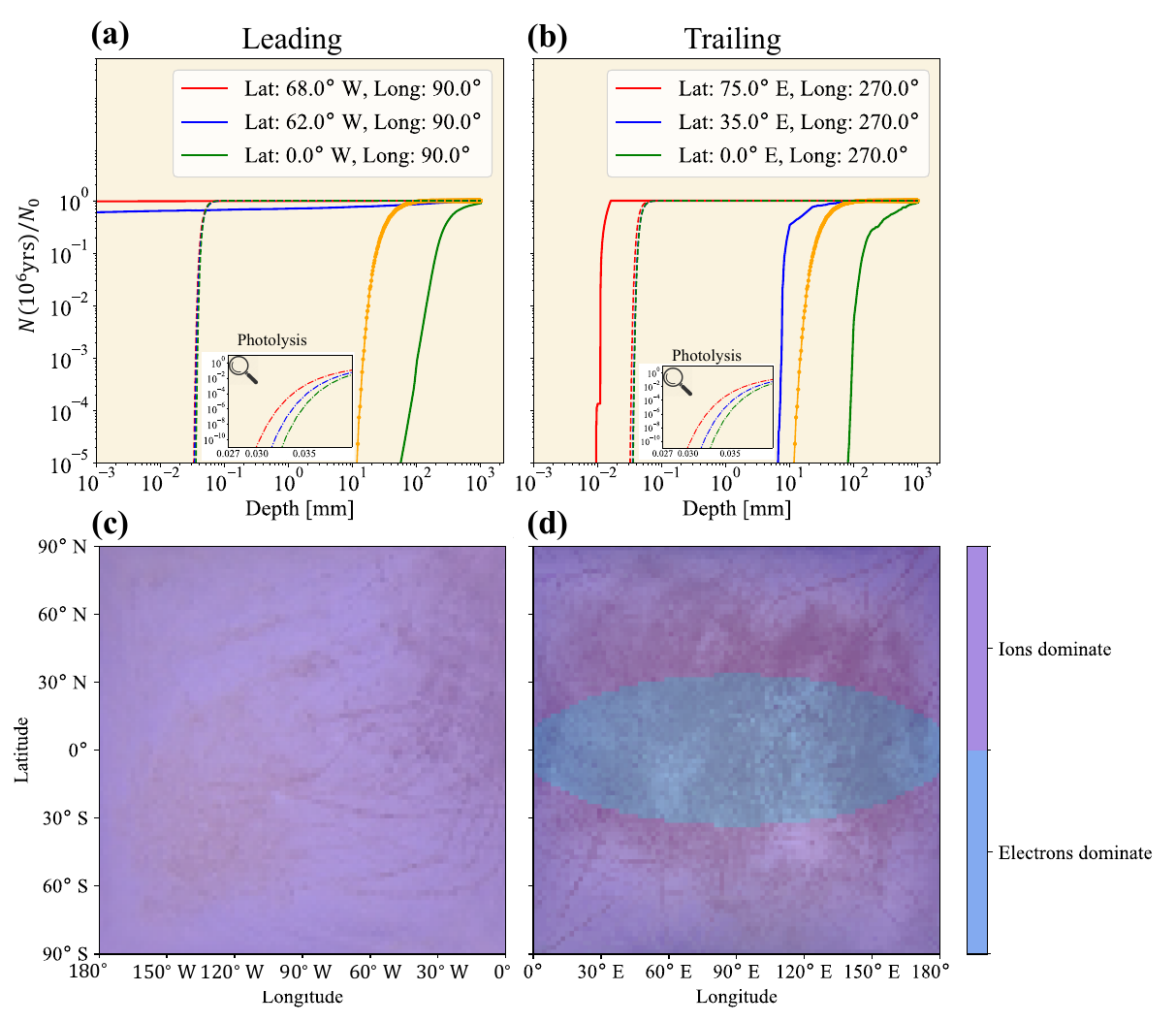}}
  \caption{Effectiveness of distinct degradation mechanisms on Europa, plotted over images of Europa's leading and trailing hemispheres taken by the Galileo spacecraft. \textbf{(a)} and \textbf{(b)} Relative amino acid concentrations at select locations on the leading and trailing hemispheres, respectively, after being subjected to each distinct and independent degradation mechanism for 10$^6$ years. \textbf{Dashed lines}: photolysis at select latitudes (colors). \textbf{Solid lines}: radiolysis by electrons at select latitudes (colors). \textbf{Dotted orange line}: radiolysis by ions, which is assumed to be uniform across Europa's surface. \textbf{(c)} and \textbf{(d)} Most effective degradation mechanism at different locations of the trailing and leading hemispheres, respectively, after 10$^6$ years, assuming a radiolytic constant of 0.034 MGy$^{-1}$. Photolysis is not shown as it is not the dominant degradation mechanism anywhere on Europa's surface.}
     \label{Fig_radiolysis}
\end{figure*}

Fig. \ref{Fig_radiolysis} demonstrates the impact of different degradation mechanisms on the relative abundances of amino acids at various locations on Europa. Given the radiolytic constant of 0.034 MGy$^{-1}$, degradation is most pronounced within the electron bombardment lenses, up to 30$^\circ$ and 60$^\circ$ for the leading and trailing hemispheres, respectively. In regions where electron bombardment is less prevalent, vapor-deposited ice significantly attenuates the penetration of UV photons, reducing photolytic degradation. As a result, in these areas, the degradation of amino acids is predominantly driven by radiolysis by ions (a detailed discussion can be found in Appendix \ref{methods_radiolysis}).

\section{Estimating Fluorescence Signal} \label{fluorescence}

We couple two degradation mechanisms to estimate the net longevity of aromatic amino acids as a function of their geographic location and depth on Europa. 
In approaches utilizing laser-induced UV spectroscopy, it is often assumed that ice is transparent \citep{eshelman2019watson}, allowing the laser to penetrate deeply and the fluorescence to propagate to the surface without severe attenuation. This assumption is inadequate for the case of Europa because, in the absence of an atmosphere to filter harmful UV photons \citep{plainaki2018towards}, transparent ice would result in the rapid photolytic degradation of amino acids to greater depths, severely limiting the effectiveness of this detection methodology.

Extinction of photon fluxes in vapor-deposited ice protects amino acids from photolytic degradation but also attenuates both the laser beam and the induced fluorescence traveling back to the surface (see Appendix \ref{methods_fluorescence}).
Our findings indicate that such a laser can effectively penetrate the uppermost millimeter of vapor-deposited ice under Europan conditions, establishing this depth as the critical longevity depth scale of amino acids for significant fluorescence signal detection (see Appendix \ref{methods_fluorescence}). 
In comparison with reported detection thresholds of organic compounds buried in at least 2.4 cm of glacial ice in Earth-like conditions \citep{eshelman2019watson}, Europa and Enceladus, to a lesser degree, present more challenging environments for applying such methodology, as the relevant layer prone for detection is also the one exposed to the harshest degradation. 

\begin{figure*}[t!]
\centering
\rotatebox[origin=c]{0}{\includegraphics[scale = 0.45]{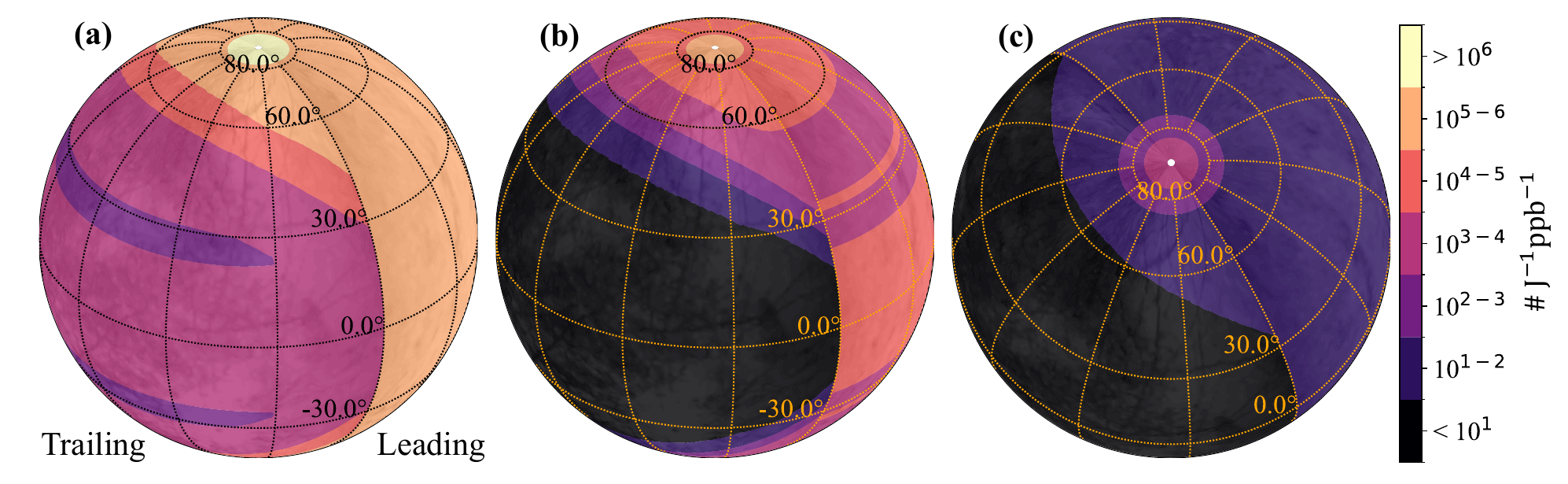}}
  \caption{Net yield of fluorescence photons per unit 248.6 nm laser energy per concentration (ppb) on Europa, given a radiolytic constant of 0.034 MGy$^{-1}$, at three exposure times to degradation mechanisms: \textbf{(a)} 10 years, \textbf{(b)} 20 years, and \textbf{(c)} 50 years. \# denotes the total number of emitted fluorescence photons.}
     \label{Fig_degradation}
\end{figure*}

To assess the detectability of aromatic amino acids through laser-induced spectroscopy, we assumed an initial concentration of tryptophan to be 0.1 ppb and a a mixing ratio of 1:36:66 for tryptophan, tyrosine, respectively. The three amino acids were modeled as uniformly-distributed across the vertical ice column. These concentrations represent conservative estimates derived from Earth's oceans \citep{yamashita2004chemical, moura2013relative}.
Fig. \ref{Fig_degradation} illustrates the net number of generated fluorescence photons, as a function of a 248.6 nm laser \citep{johnson1980physics} energy and amino acid concentration, at three nominal terrain ages, demonstrating the rapid deterioration of the potential signal due to degradation, assuming a radiolytic constant of 0.034 MGy$^{-1}$ (see Appendix \ref{methods_fluorescence}).
The structure of fluorescence yield, which serves as a proxy for the degradation rate, is primarily controlled by the effectiveness of hemispheric radiolysis lenses, particularly in the trailing hemisphere. Secondary contributions arise from uniformly bombarding ions, with a lesser influence from photolysis, which is further modulated by the ice phase.

It is evident that given a prohibitive radiolytic constant, even very low concentrations of aromatic amino acids are sufficient to generate a statistically significant fluorescence signal, given young enough terrain and potentially even for detection from orbit. 
Polar regions, which are least exposed to solar irradiance and are situated farthest from the electron radiation lenses, present the optimal locations for the detection of aromatic amino acids.
Interestingly, the limited evidence of Europa's plume activity points to the south polar region \citep{roth2014transient}. If plume activity indeed occurs, and is more likely in this region, similar to Enceladus \citep{yeoh2015understanding}, the convergence of circumstances may create favorable conditions for probing plume ejecta, enhancing the prospects for detecting aromatic amino acids.

We compute the permissible degradation of the effective upper layer of ice required to produce a statistically significant detection of aromatic amino acid fluorescence across Europa for two plausible detection strategies: A detection by an instrument at the surface or during a close flyby at a 10-kilometer altitude, for the three considered radiolytic constants (a detailed discussion can be found in Appendix \ref{methods_fluorescence}).
We show that it is feasible to detect a significant fluorescence signal within terrain exposed to Europa's surface conditions for up to several thousand years in the case of a lander mission framework and up to several hundred years in the case of a flyby mission framework, where the polar regions offer the only feasible detection margin for the latter.

\section{Conclusion} \label{conclusion}

Our study significantly advances the understanding of Europa's potential to preserve fluorescent biomolecules embedded in near-surface ice despite the harsh conditions on its surface, as well as the ability to detect these molecules. By modeling the effects of radiolysis and photolysis, we have shown that aromatic amino acids, specifically tryptophan, phenylalanine, and tyrosine, can persist within the upper millimeter of ice at high-latitude regions for hundreds of years. This finding is crucial for future astrobiological missions aiming to detect biosignatures on Europa.

The degradation rates of these biomolecules were found to vary significantly with latitude and depth, influenced by the intensity of charged particle bombardment and the phase of the ice. Our results indicate that the leading hemisphere of Europa, sparsely impacted by high-energy electrons, presents a more favorable environment for the preservation of these amino acids compared to the trailing hemisphere, which is exposed to lower-energy particles that are more numerous with shallower penetration depths.
This said, the radiolytic constant of aromatic amino acids in the near-surface ice of Europa, which plays a critical role in determining their longevity therein, is poorly constrained. Factors such as ice impurities and the manner in which organics are clumped in the ice introduce variability that necessitates further study.

We show that laser-induced fluorescence spectroscopy can effectively detect these biomolecules even from orbit. 
This technique can penetrate the uppermost millimeter of vapor-deposited ice, making it attractive for potential future missions targeting geologically young terrain, such as plume ejecta.
Specifically, our findings underscore the importance of the polar regions on Europa as prime targets for biosignature detection, where preservation of aromatic amino acids may lead to detectable concentrations of these compounds. 
Future missions equipped with spectroscopic instruments will be well-positioned to explore these regions and potentially uncover evidence of biotic or abiotic synthesis of complex biomolecules.
The ability to predict the presence of fluorescent organic molecules other than aromatic amino acids on Europa’s surface is critical for evaluating the feasibility of the UV-induced fluorescence detection approach for aromatic amino acids. Detecting these coexisting molecules can, on the one hand, attenuate the unique fluorescence signatures of the aromatic amino acids and, on the other hand, provide additional context, revealing interactions between various organics and Europa’s radiation-altered ice. Such insights could help elucidate the history and origin of the extant organic material on Europa's surface.

\section*{Acknowledgements}
We are grateful to Erin Redwing, Prof. Imke de Pater, and Dr. Chris McKay for their early insights that helped shape the direction of this work.
We are also grateful to the reviewers for their constructive reports, which helped improve the quality of the article.

\end{multicols}
\clearpage

\bibliographystyle{plainnat}
\bibliography{sample}

\clearpage

\appendix

\renewcommand{\thesection}{\Alph{section}}
\renewcommand{\thefigure}{\Alph{section}.\arabic{figure}}
\renewcommand{\thetable}{\Alph{section}.\arabic{table}}
\renewcommand{\theequation}{\Alph{section}.\arabic{equation}}
\counterwithin{figure}{section}
\counterwithin{table}{section}
\counterwithin{equation}{section}

\section{Radiolysis} \label{methods_radiolysis}

\subsection{Rationale} \label{app_radiolysis_rationale}

The radiolytic degradation of amino acids follows a first-order process similar to photolytic degradation:

\begin{equation}
C(t) = C_0 e^{-k_{\rm{radio}} D t},
\end{equation}

\noindent where $C(t)$ is the concentration of amino acids at time $t$, $C_0$ is the initial concentration, $k_{\rm{radio}}$ is the radiolytic dose coefficient, and $D$ is the deposited dose rate. To estimate the longevity of amino acids under charged particle bombardment on Europa, we utilize experimentally measured radiolytic dose coefficients determined under various regimes of ionizing radiation. For uniformly distributed phenylalanine, the radiolytic dose coefficient was found to be approximately 0.05 MGy$^{-1}$ under Europan thermal conditions. This value was derived from experimental studies involving two distinct radiolysis mechanisms: bombardment by 0.8 MeV protons \citep{gerakines2012situ} and $\gamma$-ray irradiation \citep{pavlov2024radiolytic}. 
These results indirectly relate to the findings of \citet{strazzulla1992ion}, who studied the amorphization of water ice and observed that phase transitions and corresponding reaction rates depend predominantly on the energy of the incoming radiation (within the tested range) rather than its specific type. 
The radiolytic dose coefficients of tyrosine and tryptophan were scaled to be 4.3 times smaller than that of phenylalanine. This scaling is based on $\gamma$-ray radiolysis experiments, which revealed differences in degradation timescales attributable to molecular structural variations \citep{chrysochoos1968pulse, armstrong1969pulse, cataldo2011solid}. Given a mixing ratio of 1:36:66 for tryptophan, tyrosine, and phenylalanine, respectively \citep{moura2013relative}, an average radiolytic dose coefficient of 0.034 MGy$^{-1}$ was computed using the following formula:

\begin{equation}
k_{\rm{radio}} = \frac{0.05}{4.3}\cdot\frac{1}{103} + {4.3}\cdot\frac{36}{103} + 0.05\cdot\frac{66}{103} + \frac{0.05}{4.3} \approx 0.034.
\end{equation}

\begin{figure*}[t!]
\centering
\includegraphics[scale = 0.85]{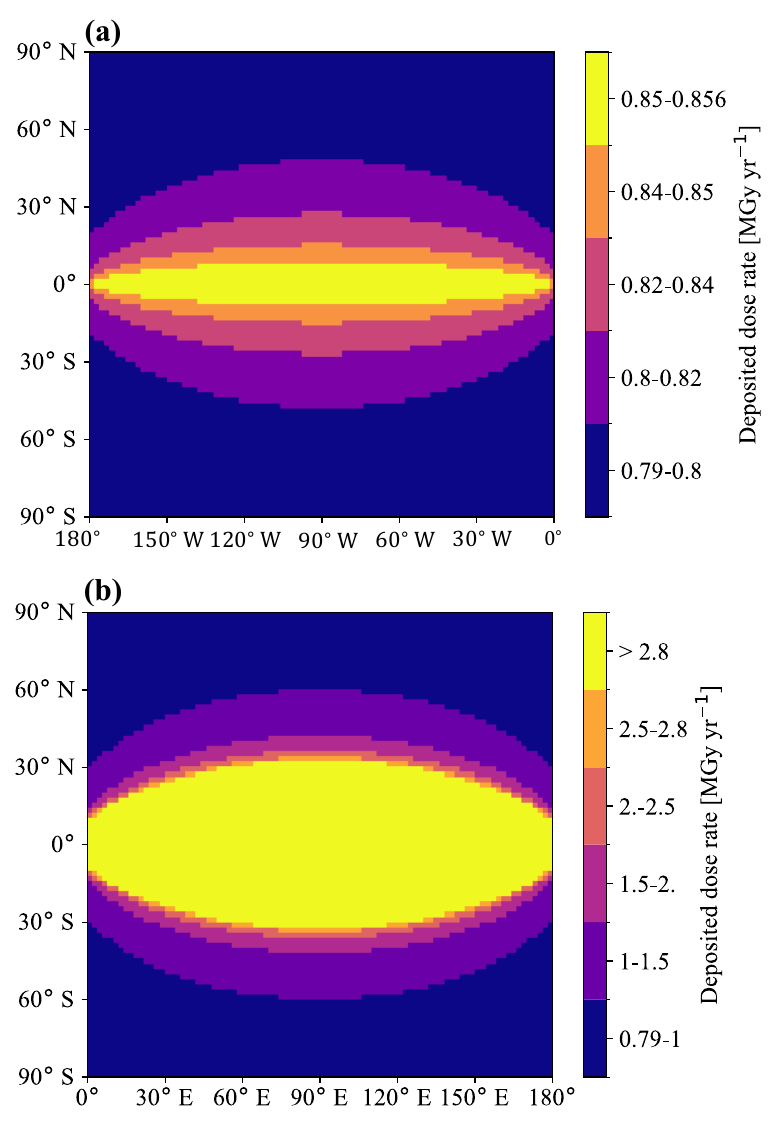}
  \caption{Integrated deposited dose rate from the bombardment of magnetospheric electrons and ions (H$^+$, O$^{2+}$, and S$^{3+}$) for the leading \textbf{(a)} and trailing \textbf{(b)} hemispheres. The bombardment of ions is assumed to be uniform, and electron bombardment patterns follow a lens-like pattern centered on the equatorial region of each hemisphere.}
     \label{fig_MGy_both}
\end{figure*}

We conducted particle transport simulations using G4beamline \citep{roberts2007g4beamline}, a \texttt{GEANT4}-based toolkit \citep{agostinelli2003geant4}, to model the depth-dependent energy deposition by energetic particles in water ice. The energy spectra for electrons and ions were derived from published fits \citep{mauk2004energetic, nordheim2018preservation, nordheim2022magnetospheric}, based on Voyager and Galileo measurements \citep{paranicas2001electron}. Depth profiles of deposited dose rates were computed for electrons, protons, and heavier ions (O$^{2+}$, S$^{3+}$) (see Figure \ref{fig_MGy_both}).

The radiolytic degradation rate at each depth and latitude was calculated using the energy deposition results from our simulations. Further details of the radiolysis simulation setup and results are provided below.

\subsection{Radiolysis Simulations} \label{app_radiolysis}

\subsubsection{Electrons} \label{app_radioysis_electrons}

For each pixel, we compute the net energy flux from all incoming electrons using the following integral:

\begin{equation}
    E(\phi,\theta) = \int_\Omega\int^{E_{\rm{max}}(\phi,\theta)}_{E_{\rm{min}}(\phi,\theta)}J_0(E)EdEd\Omega,
\end{equation} \label{eq_app_electron_KE}

\noindent where $E_{\rm{min}}$ and $E_{\rm{max}}$ are determined by their location \citep{nordheim2018preservation}. 
Following previous work \citep{paranicas2009europa}, the integral over the solid angle, accounting for downward facing flux perpendicular to the normal, equals $\pi$.


\subsubsection{G4beamline Setup} \label{app_photolysis_electrons_g4beamline}

The G4beamline simulation setup consists of two parts:

(\textbf{1}) We generate a block of 2000 detectors, each made of water ice with a density of 0.5 and 1 g cm\(^{-3}\) and dimensions \(x, y = 1 \, \text{km}\) and \(z = 0.1 \, \text{mm}\), effectively probing down to a depth of a meter. Each detector interacts with incoming electrons and records the net deposited energy. The size of the detectors is configured to ensure that the net downward energy of electrons is recorded, even if they scatter laterally over considerable distances. Additionally, we place two virtual detectors before and after the water ice detector block. These virtual detectors only count the net energy that flows through them without interacting. They account for electrons exiting the detector block on one side and backscattered electrons and $\gamma$-rays emitted by \(e^--e^+\) annihilation events, thus ensuring the conservation of energy, which should sum to the initial kinetic energy of all simulated electrons.

(\textbf{2}) The simulation of the electrons themselves: To sample correctly from the electron spectrum \citep{nordheim2018preservation}, we ensure that the region before the FWHM of the bulk of the flux (located in the lower kinetic energy limit) is sampled at least ten times. We find that the flux spectrum should be sampled logarithmically at \(\sim\)200 locations to achieve a \(<1\%\) discrepancy between numerical and analytic integration of Eq.~\eqref{eq_app_electron_KE}. At each sampled location of the spectrum, we perform a simulation involving 10$^3$ electrons at that energy interacting with the water ice detectors and correct the resulting deposited/scattered energy to the number of electrons at that energy in the flux spectrum, multiplied by \(\Delta E\) (the difference in energy between that and the next sample). The number of electrons was chosen based on output variance analysis. Following the rationale of previous work \citep{paranicas2001electron, paranicas2009europa}, the initial momentum \textit{direction} of each electron is modeled as a random variable drawn from the following distribution:

\begin{equation*}
\vec{P} =
\begin{cases}
P_x = P \sin(\theta) \cos(\phi) \\
P_y = P \sin(\theta) \sin(\phi) \\
P_z = P \cos(\theta),
\end{cases}
\end{equation*}

\noindent where \(\phi \sim \mathcal{U}(0, 2\pi)\) and \(\theta \sim \text{Cosine}(0, \pi / 2)\).

Finally, the simulation output for each pixel consists of three elements: (\textbf{1}) Depth-dependent absorbed energy within the ice detectors, (\textbf{2}) Net energy of scattered electrons that did not penetrate the ice, and (\textbf{3}) Net energy of emitted gamma photons.



\begin{figure*}[t!]
\centering
\rotatebox[origin=c]{0}{\includegraphics[scale = 0.5]{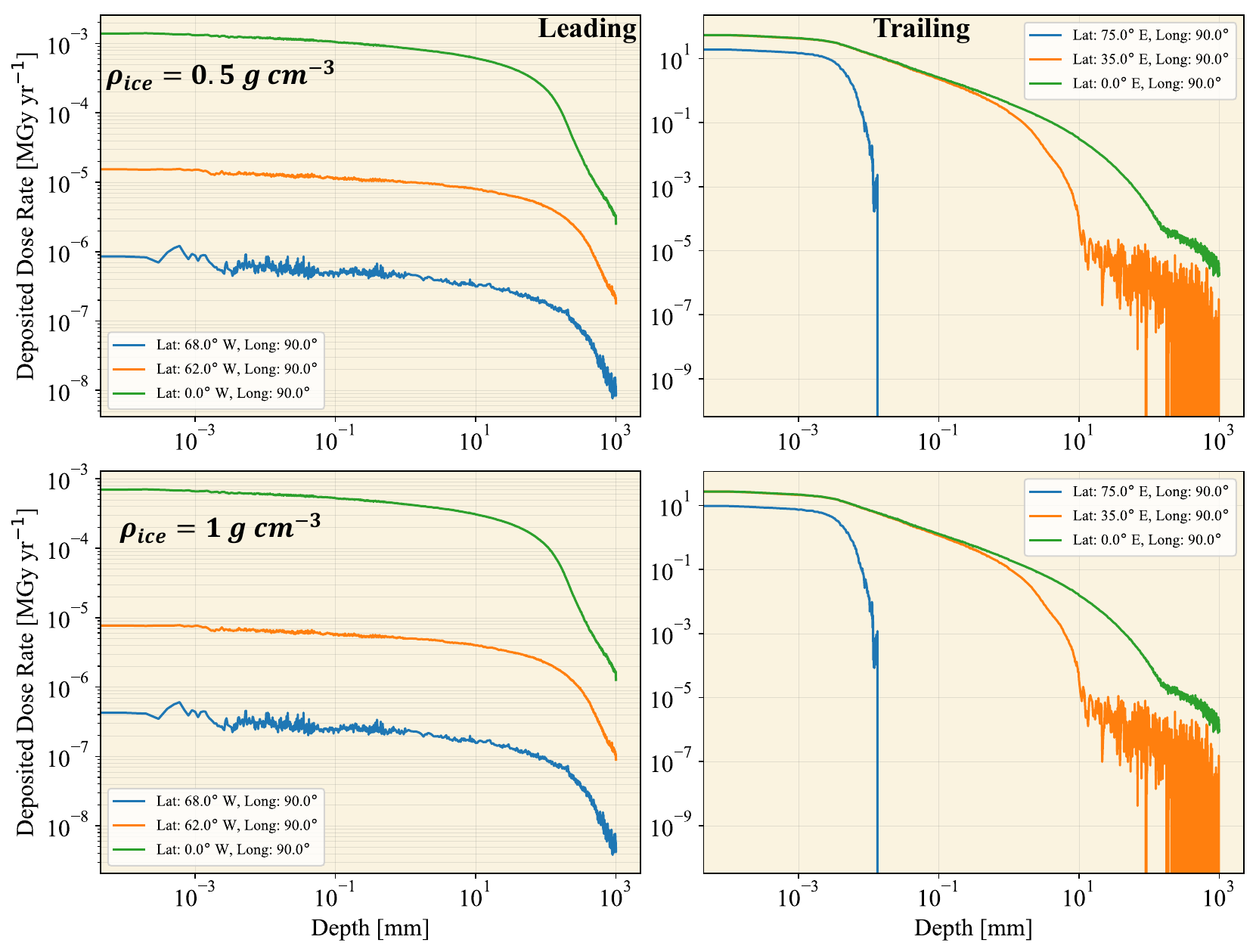}}
  \caption{Depth-wise profiles of deposited dose rate at three nominal latitudes at the leading and trailing hemispheres. \textbf{Upper Panel}: $\rho_{\rm{ice}} = 0.5$ g cm$^{-3}$. \textbf{Bottom Panel}: $\rho_{\rm{ice}} = 1$ g cm$^{-3}$.}
     \label{Fig_deposited_Mgy_example}
\end{figure*}

To convert the deposited energy in each detector from MeV to the deposited dose rate (in MGy), we divide it by $\rho \cdot dz$, where $\rho$ is the ice density and $dz$ is the thickness of each ice element.
In Fig. \ref{Fig_deposited_Mgy_example}, we plot depth-wise profiles of deposited energy flux at three nominal locations in the trailing and leading hemispheres.

\subsection{Ions} \label{app_radioysis_ions}

Recent work demonstrated that within the relevant range of energies (i.e., 0.01-100 MeV), the bombardment patterns of heavy ions follow a uniform trend as the kinetic energy of the ion increases \citep{nordheim2022magnetospheric}, providing also flux spectra for the three energetic ions, which we adopt for this work.

\begin{figure*}[t!]
\centering
\rotatebox[origin=c]{0}{\includegraphics[scale = 0.5]{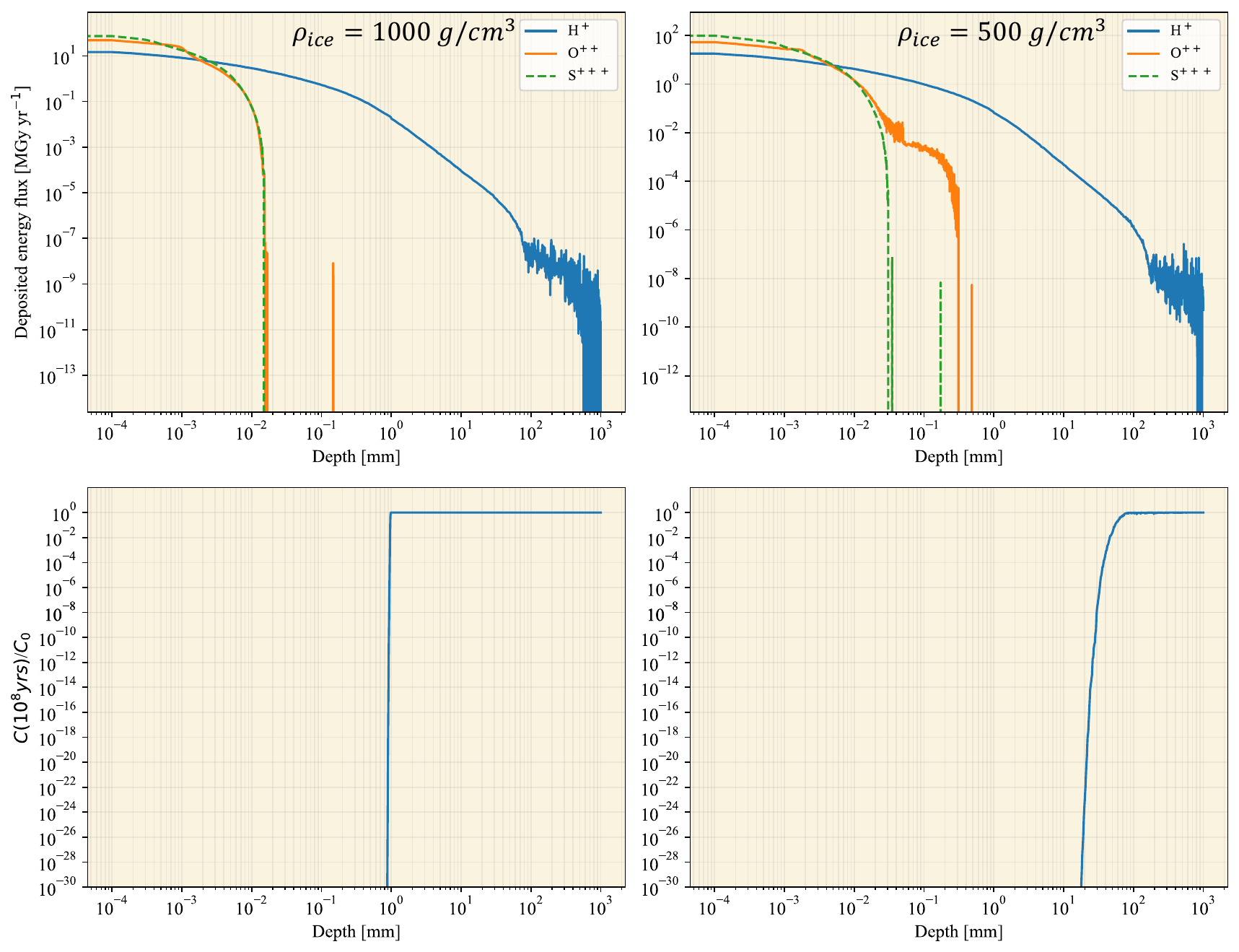}}
  \caption{\textbf{Top Panel}: Energy deposition flux of the three major magnetospheric ions as a function of ice depth at different ice densities. \textbf{Bottom Panel}: relative amino acids concentration as a function of depth after 10$^8$ years of being subjected to heavy ion bombardment.}
     \label{Fig_deposited_Mgy_ions}
\end{figure*}

In Fig. \ref{Fig_deposited_Mgy_ions}, we plot depth profiles of the deposited dose rates by the three ions, as well as the fractional concentration of amino acids as a function of depth after having been exposed to heavy ion bombardment for 10$^8$ years. 

\section{Photolysis} \label{methods_photolysis}

The photolytic degradation of amino acids follows a first-order decay law \citep{ehrenfreund2001photostability, johnson2012ultraviolet}:

\begin{equation} \label{eq_photolysis}
C(t) = C_0 e^{-k_{\rm{photo}} t},
\end{equation}

\noindent where \(C(t)\) is the amino acid concentration at time \(t\), \(C_0\) is the initial concentration, and \(k_{\rm{photo}}\) is the photolytic reaction rate, which is both flux- and wavelength-dependent \citep{johnson2012ultraviolet}. To parameterize \(k_{\rm{photo}}\) across the spectral range, we define it as:

\begin{equation}
k(\lambda) = n(\lambda)\lambda \alpha(\lambda),
\end{equation}

\noindent where \(\lambda\) is the wavelength, \(n(\lambda)\) is the solar photon flux at wavelength \(\lambda\), and \(\alpha(\lambda)\) is a coefficient relating photon energy and flux to the photolytic reaction rate. We derive \(\alpha(\lambda)\) by inverting the half-life times reported in \citet{johnson2012ultraviolet} in their Table 2, according to:

\begin{equation}
\alpha(\lambda) = \frac{\ln(2)}{t_{1/2}} \cdot \frac{1}{n(\lambda)\lambda},
\end{equation}

\noindent where \(t_{1/2}\) is the measured photolytic half-life time. These values, measured at specific wavelengths, are extrapolated using known absorption curves for aromatic amino acids \citep{beaven1952ultraviolet}, from which wavelength-dependent reaction rates are derived. The extrapolation of \(\alpha(\lambda)\) across the spectral range is detailed in Appendix \ref{app_photolysis_alpha}.

The incident solar flux at different latitudes is given by:

\begin{equation}
    F(\phi, \lambda) = S_0(\lambda)f(\phi)(1 - P)(1 - \alpha),
\end{equation} \label{eq_average_flux}

\noindent where \(\phi\) is the latitude, \(f(\phi)\) is the fraction of solar flux projected onto the surface at latitude \(\phi\), \(P = 0.033\) represents the fraction of Europa’s time in Jupiter's shadow, and \(\alpha \approx 0.68\) is Europa’s disk-averaged Bond albedo \citep{grundy2007new}. The function \(f(\phi)\) is derived in \citet{ashkenazy2019surface}.

The depth-dependent attenuation of radiation in the ice is modeled using the Beer-Lambert law:

\begin{equation}
    F(\phi,\lambda, z) = F(\phi, \lambda) e^{-\mu(\lambda) z},
\end{equation} \label{eq_beer_lambert}

\noindent where \(F(\phi,\lambda)\) is the latitudinal and wavelength-dependent flux at the surface, \(F(\phi,\lambda, z)\) is the flux at depth \(z\), and \(\mu(\lambda)\) is the wavelength-dependent extinction coefficient of ice.

The depth-dependent photolytic reaction rate at each latitude is calculated as a weighted average across the spectral range:

\begin{equation}
    k_{\rm{photo}, i}(\phi, z) = \sum_{j \in [\lambda_{\rm{min}}, \lambda_{\rm{max}}]}k_{\rm{photo},i}(\phi, j, z)\frac{n_j}{N},
\end{equation} \label{eq_k_wighted_avg}

\noindent where \(N\) is the integrated fluence across the spectral range and \(n_j\) is the fluence at the \(j\)th wavelength.

\subsection{Extrapolating the Wavelength-Dependent Reaction Coefficients} \label{app_photolysis_alpha}

\begin{figure*}
\centering
\rotatebox[origin=c]{0}{\includegraphics[scale = 0.8]{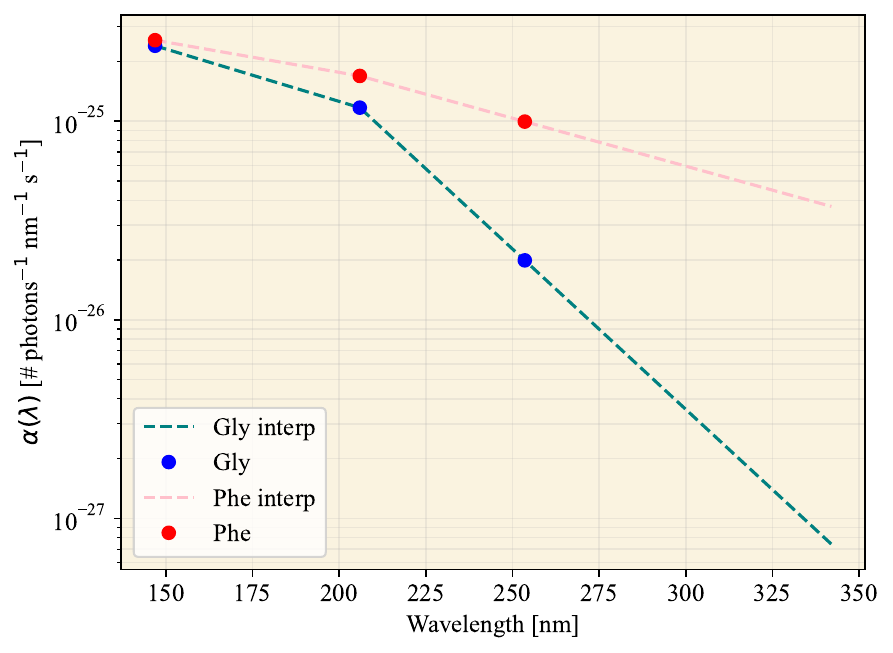}}
\caption{Wavelength-dependent reaction coefficient \(\alpha(\lambda)\), measured and extrapolated. \textbf{Dots}: \(\alpha(\lambda)\) inverted from half-life times measured by \cite{johnson2012ultraviolet}. \textbf{Dashed lines}: Linear interpolation of the inverted \(\alpha(\lambda)\) for the 147-342 nm range.}
\label{Fig_alpha_interp}
\end{figure*}

 We interpolate \(\alpha(\lambda)\) across the 147-342 nm range using linear interpolation of \(\ln(\alpha(\lambda))\), ensuring a gradual transition of \(\alpha(\lambda)\) as the wavelength increases. This approach is conservative, given the substantial decrease in absorption for wavelengths \(\lambda > \sim 300\) nm \citep{fodor1989deep, wen1998uv, orzechowska2007ultraviolet}. The extrapolated \(\alpha(\lambda)\) values are shown in Fig. \ref{Fig_alpha_interp}.
Lastly, the photolytic degradation rates of tryptophan and tyrosine were scaled by 4.3, based on the rationale presented in Appendix \ref{app_radiolysis_rationale}.

\subsection{Spectral Range} \label{app_photolysis_spectral_range}

Decarboxylation, the process involving the removal of a carboxyl group (-COOH) from a molecule, was demonstrated to be a notably favored chemical transformation of amino acids under UV irradiation \citep{lion1981spin, ehrenfreund2001photostability}. The preeminence of decarboxylation as the primary photolytic pathway for aromatic amino acids can be attributed to the structural configuration of the carboxyl group within these molecules. The proximity of the carboxyl group to the aromatic ring facilitates facile $C_{\alpha}-C_{\beta}$ (where $C_{\alpha}$ is the carbon adjacent to the carboxyl group and $C_{\beta}$ is the carbon within the side chain) bond cleavage upon photon absorption, yielding a carboxyl radical intermediate \citep{tseng2010photostability}. This intermediate species undergoes subsequent rearrangement and fragmentation processes, culminating in the release of carbon dioxide (CO$_2$) and the generation of a stable radical or carbonation \citep{ehrenfreund2001photostability, rahman2019recent}.

Photolytic degradation of amino acids on Europa’s surface was experimentally explored in previous work \citep{orzechowska2007ultraviolet, johnson2012ultraviolet}, which investigated the half-life times of glycine and phenylalanine under the influence of solar irradiation at wavelengths of 147, 206, and 254 nm. Since these wavelengths don’t fully encapsulate the effective photolytic degradation range, we opted for a broader spectral range for analysis, specifically from 147 to 342 nm.

The choice of 342 nm as the lower limit is justified by its equivalence to the $C_{\alpha}-C_{\beta}$ bond dissociation energy in glycine ($\approx$ 3.62 eV \citep{luo2002handbook}). In aromatic amino acids, resonance stabilization due to the presence of an aromatic ring enhances the strength of adjacent bonds \citep{aihara1992aromatic}. This stabilization, facilitated by the delocalization of $\pi$ electrons, notably reinforces the $C_{\alpha}-C_{\beta}$ bond. Conversely, glycine lacks aromatic rings, resulting in a lower dissociation energy for its $C_{\alpha}-C_{\beta}$ bond. Hence, adopting 342 nm as the lower limit sets a conservative range for considering wavelengths in the photolytic process.

While previous studies have documented the photolytic degradation of amino acids at wavelengths as short as 121.6 nm, corresponding to the Lyman-$\alpha$ peak in the solar spectrum \citep{ehrenfreund2001photostability}, water ice, regardless of whether it's crystalline or amorphous, effectively attenuates photons below approximately 150 nm \citep{he2022refractive}. 
Considering that (\textbf{1}) our investigation primarily focuses on amino acids within the ice matrix, and (\textbf{2}) half-life times of amino acids found on the surface are very short regardless of whether shorter wavelengths are considered, we set 147 nm -- the shortest wavelength analyzed in previous experimental work \citep{johnson2012ultraviolet} -- as the upper energetic threshold for our analysis.

\section{Thermal History of Vapor-Deposited Ice} \label{methods_ice}

\begin{figure*}[t!]
\centering
\rotatebox[origin=c]{0}{\includegraphics[scale = 0.8]{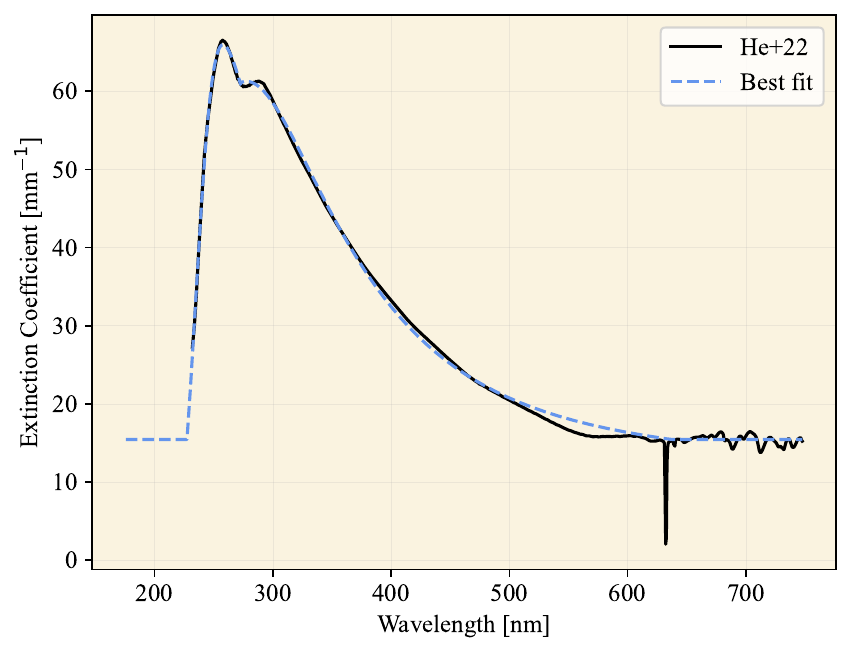}}
\caption{Extinction coefficient profile fitting procedure demonstration. Solid black line: measured extinction coefficient spectrum for vapor-deposited ice at 100 K. Dashed blue line: Best fit curve.}
\label{Fig_fitting_mu}
\end{figure*}

At different deposition temperatures, we fit the extinction spectra using a function composed of two skewed Lorentzians and a left tail, defined as the average of the right tail for \(\lambda > 600\) nm (see \ref{Fig_fitting_mu}). The resulting fit functions are concatenated and interpolated over the temperature range of 30–160 K.

Experimental work has shown that vapor-deposited amorphous ice, when deposited below 110 K and heated, exhibits significant changes in optical properties \citep{he2022refractive}. In contrast, the optical properties of crystalline ice, which doesn't undergo additional significant thermal annealing, remain stable \citep{jenniskens1996crystallization, mitchell2017porosity, he2022refractive}. To account for this, we compute the diurnal surface temperature cycle at each latitude and calculate the average change in optical properties due to post-depositional heating. The thermal evolution is modeled within the top $\sim$5 cm of the surface, assuming a uniform ice density as supported by gravitational compaction studies on Europa \citep{mergny2024gravity}.

The adjustment of the extinction coefficient profile due to thermal evolution is performed as follows:

\begin{itemize}
    \item The diurnal surface temperature cycle is computed for each latitude by integrating a partial differential equation of the diurnal temperature cycle as a function of depth, derived and specified with appropriate boundary conditions and parameter values in previous work \citep{ashkenazy2019surface}. We integrated the equation for about fifty days to capture the diurnal heating effects within the top few centimeters.
    \item We interpolate the thermal evolution of the extinction coefficient at 500 nm based on experimental results \citep{he2022refractive}, using the curves of the slower heating rates (0.6 K/min) that best reflect diurnal rates on Europa’s surface \citep{ashkenazy2019surface}.
    At each latitude, we assume that the extinction coefficient is heated from the average- to the maximal- diurnal temperature (i.e., averaging over all possible deposition phases of the diurnal cycle).
    \item The average thermally-evolved extinction coefficient is computed at each latitude by scaling the extinction coefficient spectrum by the ratio of thermally-evolved and thermally unchanged values at 500 nm, assuming that the spectral absorption curve (e.g., Fig. \ref{Fig_fitting_mu}) remains constant.
\end{itemize}

The derived extinction coefficient profile accounts for the varying thermal conditions across latitudes, incorporating both amorphous and crystalline ice properties.

\subsection{Extinction Coefficients of Vapor-Deposited Ice} \label{app_thermal}

Vapor-deposited ice is of particular interest due to three factors:

\begin{trivlist}
\item \textbf{Prevalent in Higher Latitudes}: Previous work suggested that the surface ice on Europa’s leading hemisphere follows a bi-modal regime, with amorphous ice prevalent in high latitudes ($\phi > 60^\circ$) \citep{berdis2020europa}. This is due to slower crystallization driven by diurnal temperature cycles \citep{jenniskens1996crystallization}. Beyond latitude 75$^\circ$, this process is even slower than the estimated upper limit of the mean turnover rate of Europa’s surface \citep{ip2000magnetospheric, bierhaus2005secondary, doggett2009geologic}. In contrast, lower latitudes are characterized by predominantly crystalline ice. Simulations indicate that most charged particle energy is deposited at lower latitudes, potentially degrading embedded amino acids to greater depths. Additionally, average solar irradiance decreases with increasing latitude, reducing photolytic degradation potential \citep{ashkenazy2019surface}.

\item \textbf{Indicative of Geologically-active Terrain}: Plume ejecta is suggested to be primarily deposited as vapor-deposited ice \citep{goodman2004hydrothermal, quick2020characterizing}. Although the evidence is sparse, plume activity on Europa (and Enceladus) has been traced to the south pole region \citep{roth2014transient}.

\item \textbf{Strong Scattering Properties}: The refractive indices and extinction coefficients of vapor-deposited water ice were recently measured at different temperatures \citep{he2022refractive}, distinguishing two steady-state regimes: amorphous ice at $T\lesssim 130$ K and crystalline ice at $T\gtrsim 160$ K. The significant difference between the mean path length for UV-vis absorption (hundreds of meters) \citep{warren2008optical} and scattering (tens of microns) \citep{he2022refractive} indicates that vapor-deposited ice can effectively attenuate incoming solar radiation, prolonging the lifetime of embedded molecules.

Previous works demonstrated how optical properties of water ice change with different composition, deposition, and thermal evolution mechanisms \citep{warren2008optical, warren2019optical}. For example, measured photolytic lifetimes of amino acids in crystalline ice indicated that the mean longevity of the considered species under Europan conditions is on the order of a decade \citep{orzechowska2007ultraviolet}. In contrast, the optical properties of vapor-deposited ice, particularly its scattering coefficient, differ significantly from crystalline ice \citep{he2022refractive}. Therefore, lifetime estimates for amino acids in Europan near-surface ice can vary considerably depending on the ice’s properties.

\end{trivlist}

\begin{figure*}[t!]
\centering
\rotatebox[origin=c]{0}{\includegraphics[scale = 0.8]{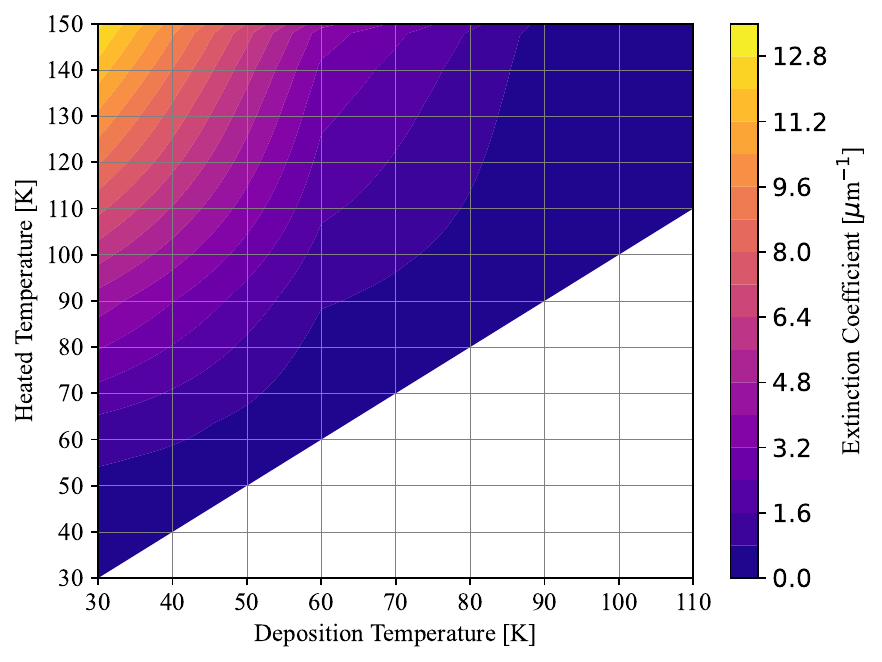}}
\caption{Thermal history of vapor-deposited amorphous ice as a function of deposited and heated temperatures, as reported in \citet{he2022refractive} (their Fig. 13). The heating gradients are 0.6 K/min, except for ice deposited at 30 K, for which the gradient was 3 K/min.}
\label{Fig_thermal_history_he}
\end{figure*}

\section{Amorphization and Crystallization} \label{app_amorphization_crystallization}

The amorphization of ice is dependent on both temperature and absorbed radiation dose, measured in previous experimental work \citep{strazzulla1992ion}. We estimate the mean temperature of near-surface ice over multi-annual timescales by solving the partial differential equation of the diurnal cycle temperature cycle as a function of depth \citep{ashkenazy2019surface}, until the diurnal cycle reaches steady-state. The deposited dose is calculated using radiolysis simulations, converting the absorbed dose from MGy to eV/H$_2$O.

For crystallization, we solve the Avrami equation (with \(n = 2\) and \(\Delta H = 60 \, \rm{kJ/mol}\) based on prior measurements \citep{jenniskens1996crystallization, mitchell2017porosity}), which describes the bulk of the crystallization process of porous ice. The net crystallinity at any given location and time is derived by balancing the amorphization and crystallization rates.

The final extinction coefficient for vapor-deposited ice is then computed as a combination of thermally evolved amorphous ice and crystalline ice:

\begin{equation}
    \mu_{\rm{final}}(\lambda) = (1 - f_{\rm{total}})\cdot\mu_{\rm{thermal}}(\lambda) + f_{\rm{total}}\cdot\mu_{\rm{crystalline}}(\lambda),
\end{equation}

\noindent where \(f_{\rm{total}}\) is the net crystalline fraction, and \(\mu_{\rm{thermal}}\) and \(\mu_{\rm{crystalline}}\) represent the extinction coefficients of thermally evolved and crystalline ice (i.e., that deposited at 160 K \citep{he2022refractive}), respectively.
It is important to remark that crystalline ice that was impacted by radiat§ion (i.e., amorphized) and ice that was deposited as amorphous ice do not necessarily share the same optical properties \citep{berdis2020europa}. This simplifying assumption is made here because these two cases have hitherto not been disambiguated in terms of their resulting optical (and other) properties.

The net crystallization fraction at each location, depth, and time, is computed as follows:

\begin{equation}
    f_{\rm{total}}(T, D, t) = \rm{max}[0, f_{\rm{crystalline}}(T, t) - f_{\rm{amorphous}}(T, D,t)],
\end{equation} \label{eq_fraction_crystalline}

\noindent where $T$ is the depth-dependent temperature, $D$ is the depth-dependent deposited dose rate, $t$ is time, $f_{\rm{total}}$ is the net crystalline fraction, $f_{\rm{crystalline}}$ is the crystalline fraction computed with the Avrami equation, and $f_{\rm{amorphous}}$ is the amorphous fraction derived from experimental measurements by irradiation of ice by $\gamma$-rays \citep{strazzulla1992ion}. There, it was also discussed that the amorphization of ice due to irradiation is rather dependent on the kinetic energy of the radiation than on its type.

\begin{figure*}[t!]
\centering
\rotatebox[origin=c]{0}{\includegraphics[scale = 1.]{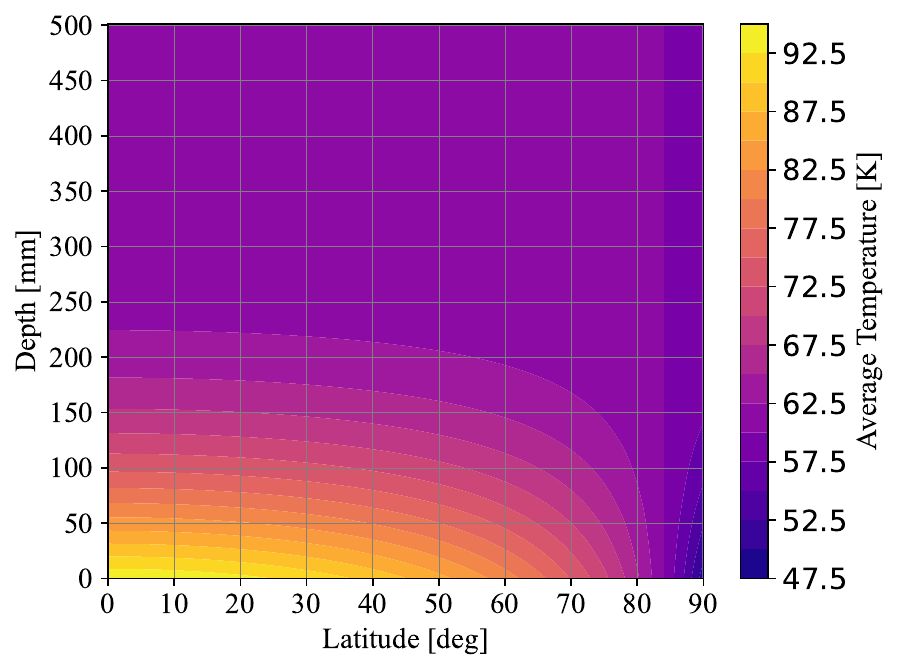}}
\caption{Average diurnal temperature of near-surface Europan ice as a function of latitude and depth.}
\label{Fig_avg_T_z}
\end{figure*}

In Fig. \ref{Fig_avg_T_z}, we plot the average diurnal temperature of near-surface ice on Europa as a function of latitude and depth, computed by integrating a partial differential equation of the diurnal temperature cycle as a function of depth, derived and specified with appropriate boundary conditions and parameter values in previous work \citep{ashkenazy2019surface}.

\begin{figure*}[t!]
\centering
\rotatebox[origin=c]{0}{\includegraphics[scale = 0.5]{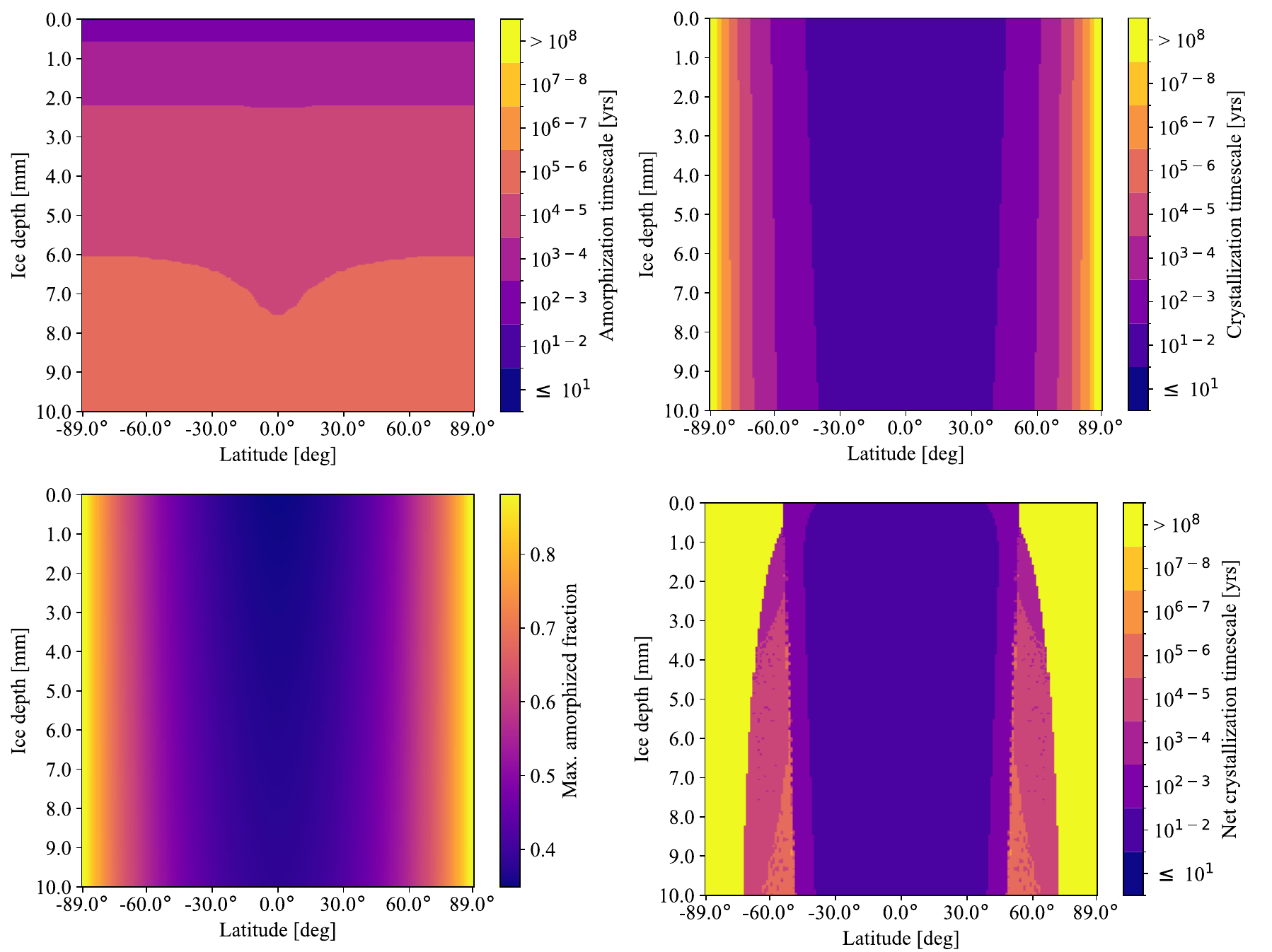}}
\caption{Amorphization and Crystallization timescales for the leading hemisphere at the prime meridian. \textbf{Upper Panel}: Amorphization and Crystallization timescales, respectively. The amorphization timescale is set for either a relative fraction of 0.5, where the temperature is sufficiently low, or for the maximum fraction otherwise. \textbf{Bottom Left Panel}: Maximum achievable amorphization fraction as a function of latitude. \textbf{Bottom Right Panel}: Net crystallization timescale, accounting for both amorphization and crystallization.}
\label{Fig_amor_crys_leading}
\end{figure*}

\begin{figure*}[t!]
\centering
\rotatebox[origin=c]{0}{\includegraphics[scale = 0.5]{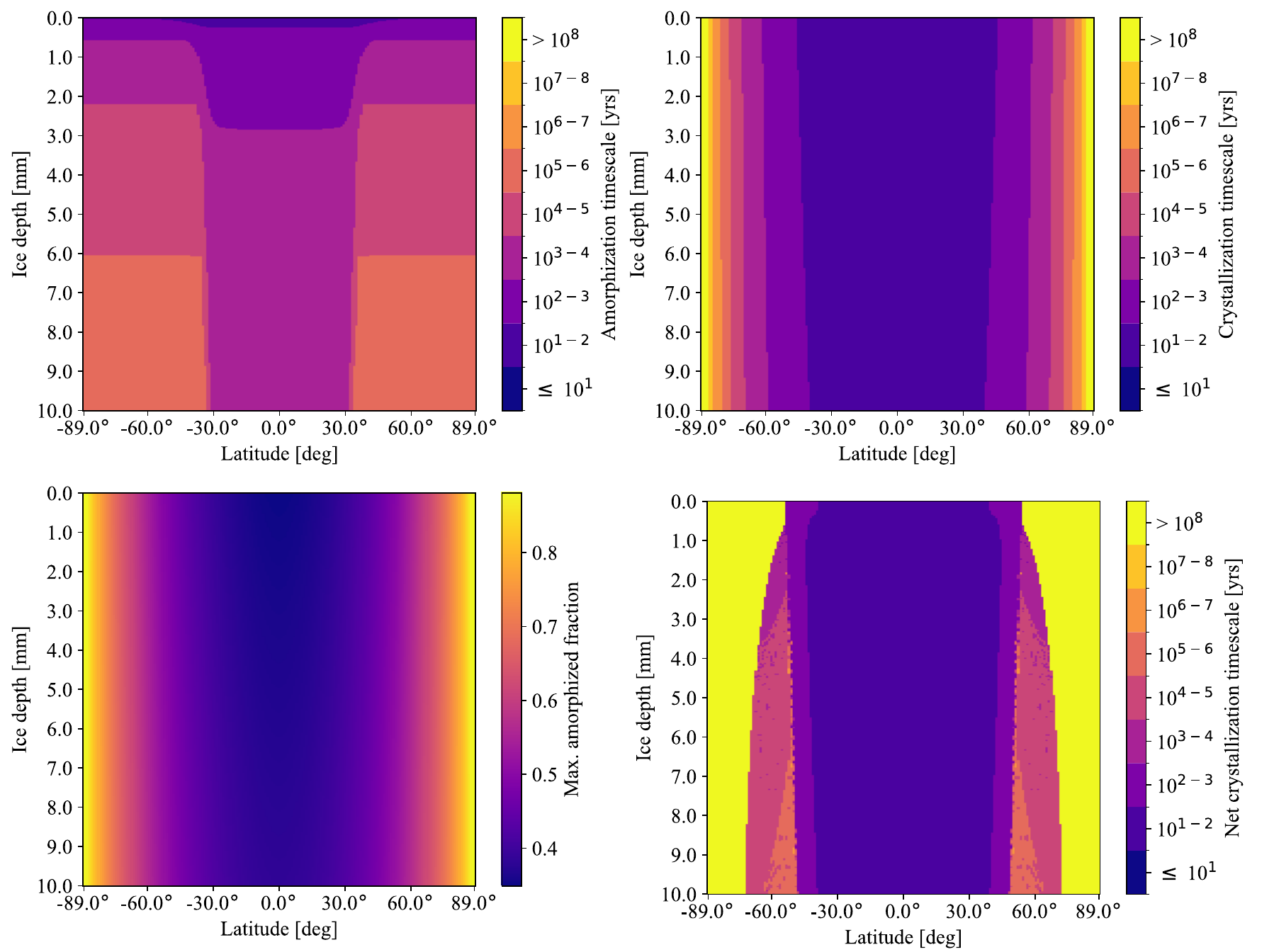}}
\caption{Amorphization and Crystallization timescales for the trailing hemisphere at the antimeridian, similar to \ref{Fig_amor_crys_leading}.}
\label{Fig_amor_crys_trailing}
\end{figure*}

In Figs. \ref{Fig_amor_crys_leading} and \ref{Fig_amor_crys_trailing}, we plot the computed timescales of crystallization and amorphization of near-surface ice as a function of latitude for the leading and trailing hemispheres, respectively.

\section{Fluorescence} \label{methods_fluorescence}

The total concentration of amino acids as a function of depth and time, accounting for both photolytic and radiolytic degradation, is given by:

\begin{equation}
M(\phi, \theta, z, t) = \sum_{i \in [\rm{phe}, \rm{tyr}, \rm{trp}]} M_i(\phi, \theta, z, t=0) e^{-(k_{\rm{photo},i}(\phi, z) + k_{\rm{radio},i} D(\phi, \theta, z)) t},
\end{equation}

\noindent where \(M_i\) is the molarity of the $i$th amino acid species, \(\phi\) is the latitude, \(\theta\) is the longitude, \(k_{\rm{photo}}\) is the photolytic reaction rate, and \(k_{\rm{radio}}\) is the radiolytic constant.  We compute the initial molarity as follows:

\begin{equation}
M_i = \frac{p_{\rm{amino}}}{p_{\rm{H_{2}O}}}\cdot \frac{\rho_{\rm{\rm{ice}}}}{M_{\rm{H_{2}O}}},
\end{equation}

\noindent where $p_{\rm{amino}}$ is parts amino acids of the $i$th species, $p_{\rm{H_{2}O}}$ is parts water (where e.g., for 1 ppb, $\frac{p_{\rm{amino}}}{p_{\rm{H_{2}O}}} = \frac{1}{10^9}$), $\rho_{\rm{\rm{ice}}}$ is the ice density, and $M_{\rm{H_{2}O}} \approx 18.016$ g mol$^{-1}$ is the molar mass of H$_2$O. The attenuation of the laser beam as it penetrates the ice is modeled using:

\begin{equation}
    I(z) = I_0 \alpha e^{-\left(\mu + \sum_{i=1}^{3} \epsilon_i M_i(z)\right) z},
\end{equation}

\noindent where $I(z)$ is the intensity of the laser beam at depth $z$, $I_0$ is the initial beam intensity, $\alpha$ is the albedo, $\mu$ is the extinction coefficient of the ice, $\epsilon_i$ is the molar absorptivity of the $i$th amino acid species, and $M_i(z)$ is the concentration of the $i$th amino acid species at depth $z$.

At each depth, the total energy converted into fluorescence for each amino acid species within an infinitesimal depth interval $dz$ is given by:

\begin{equation}
    \int_{-\infty}^{\infty} I_{\rm{fluo},i}(\lambda, z) d\lambda = I(z) \Phi_i \epsilon_i M_i(z) dz,
\end{equation}

\noindent where \(I_{\rm{fluo},i}(\lambda, z)\) is the characteristic fluorescence spectrum for the $i$th amino acid species, and $\Phi_i$ is the fluorescence quantum yield of that species. The fluorescence-related properties of the aromatic amino acids are listed in Table \ref{tab_amino_fluo_properties}.

\begin{table}[t!]
\footnotesize
    \centering
    \begin{tabular}{|c|c|c|}
    \hline\hline
        Species & Quantum Yield ($\Phi$) & Molar Absorptivity ($\epsilon$) [cm$^{-1}$ mol$^{-1}$ L] \\
         \hline
         tryptophan & 0.12 & 5579 \\
         tyrosine & 0.13 & 1140 \\
         phenylalanine & 0.022 & 88 \\
         \hline
    \end{tabular}
    \caption{Fluorescence properties of aromatic amino acid species. 
    Quantum yield and molar absorptivity values were adopted from published experimental work \citep{chen1972measurements,fasman1975handbook}.}
    \label{tab_amino_fluo_properties}
\end{table}

Fluorescence emitted from depth $z$ propagates back to the surface, attenuated by the extinction coefficient of the ice. The intensity at the surface is given by:

\begin{equation}
I_{\rm{surface},i}(\lambda) = I_{\rm{fluo},i}(\lambda, z) e^{-\mu(\lambda) z}.
\end{equation}

\noindent In Appendix \ref{app_flux_to_photons}, we describe how the emergent fluorescence signal is converted into a signal-to-noise ratio for potential detection by an orbital or surface-based instrument.

\subsection{Fluorescence Emission Attenuation and Potential Detection Strategies} \label{app_flux_to_photons}

To account for the amount of photons subtended by the telescope, assuming it is a Lambertian point-source emitter, we compute the fraction of the solid angle of the telescope, given by:

\begin{equation}
\frac{\Omega_{\rm{telescope}}}{\Omega_{\rm{total}}} = \frac{\pi D_{\rm{telescope}}^2}{4}\frac{1}{4\pi(d+z)^2} \approx \frac{D^2}{16d^2},
\end{equation}

where $D_{\rm{telescope}}$ is the diameter of the collecting area and $d$ is its distance from the surface.

\begin{figure*}[t!]
\centering
\rotatebox[origin=c]{0}{\includegraphics[scale = 0.45]{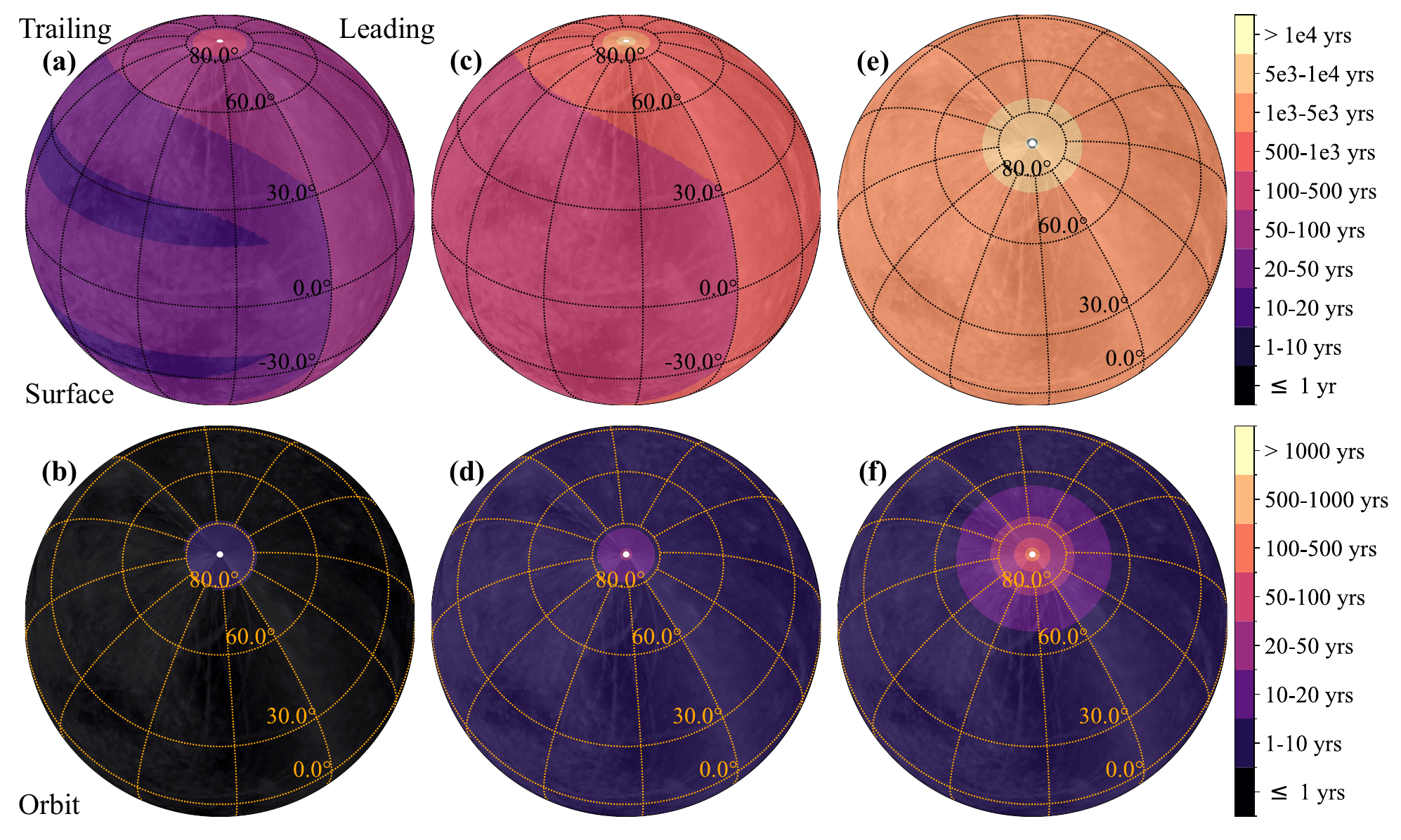}}
  \caption{Degradation timescales of aromatic amino acids within vapor-deposited ice, allowing for an SNR of 3 detection of laser-induced fluorescence, for three radiolytic constants $k_{\rm{radio}}$: (\textbf{a}, \textbf{b}) 0.034 MGy$^{-1}$ (\textbf{c}, \textbf{d}) 0.0034 MGy$^{-1}$(\textbf{e}, \textbf{f}) 0.00034 MGy$^{-1}$.
  \textbf{Top}: Detection from the surface. \textbf{Bottom}: Detection from orbit.}
     \label{fig_fluorescence_ages}
\end{figure*}

Several studies have explored the optimal approach to UV spectroscopy of amino acids, particularly for Martian Rover missions \citep{beaven1952ultraviolet, eshelman2018detecting, eshelman2019watson, lymer2021uv}. The 266 nm Nd:YAG, or the 248.6 nm krypton fluoride (KrF) excimer laser systems \citep{geusic1964laser, johnson1980physics} are usually considered as the prime candidates for space-based laser-induced UV-vis spectroscopy \citep{beegle2015sherloc}. However, this wavelength coincides with high extinction in vapor-deposited ice, leading to significant beam attenuation. A shorter wavelength around 225 nm is preferable, as it operates in a more transparent spectral region of ice, reducing attenuation and enhancing detection \citep{he2022refractive}. 
That said, it has not been examined whether such small-wavelength laser can effectively induce fluorescence of aromatic amino acids, rather than destroying them.
In Fig. \ref{fig_fluorescence_ages}, we show the extent of permissible degradation to produce a detectable fluorescence signal (SNR) of three, given the two hypothesized probing scenarios and three considered radiolytic constant values.

\begin{table}[t!]
    \centering
    \begin{tabular}{|c|c|c|}
    \hline\hline
        Parameter & Surface & Orbit \\
         \hline
         Laser beam energy [J] & 10 & 100 \\
         Collecting area diameter [m] & 0.011 & 0.5 \\
         Distance from the surface [m] & 0.048 & 10$^4$ \\
         \hline
    \end{tabular}
    \caption{Instrumental setups considered for calculation of fluorescence signal significance for the surface and orbital scenarios, respectively.
    For the surface setup, the collecting area diameter and distance from the surface are taken from the SHERLOC instrument manual \citep{beegle2015sherloc}.}
    \label{tab_instrumental_setups}
\end{table}

\end{document}